\documentclass[aps,prx,twocolumn,showpacs,floatfix,nobibnotes,nofootinbib,superscriptaddress,amsmath,amssymb,longbibliography]{revtex4-1}
\usepackage[english]{babel}
\usepackage[colorlinks,urlcolor={blue}]{hyperref}
\usepackage{eurosym}
\usepackage[usenames]{color}
\usepackage{graphicx}
\usepackage{amsmath}
\usepackage{amsfonts}
\usepackage{amssymb}
\usepackage{mathrsfs}
\usepackage{bm}
\usepackage{verbatim} 
\usepackage{ulem}
\usepackage{siunitx}
\usepackage{upgreek}

\usepackage[textsize=small]{todonotes}

\newcommand{\beq}{\begin{equation}}
\newcommand{\eeq}{\end{equation}}
\newcommand{\bea}{\begin{eqnarray}}
\newcommand{\eea}{\end{eqnarray}}

\begin{document}

\title{A new scheme for isomer pumping and depletion with high-power lasers}
\author{C.-J.~Yang}
\affiliation{ELI-NP, ``Horia Hulubei" National Institute for Physics and Nuclear Engineering, 30 Reactorului Street, RO-077125, Bucharest-Magurele, Romania}
\email{chieh.jen@eli-np.ro (corresponding author)}
\author{K.~M. Spohr}
\affiliation{ELI-NP, ``Horia Hulubei" National Institute for Physics and Nuclear Engineering, 30 Reactorului Street, RO-077125, Bucharest-Magurele, Romania}
\affiliation{School of Computing, Engineering and Physical Sciences, University of the West of Scotland, High Street, PA1 2BE, Paisley, Scotland}
\email{klaus.spohr@eli-np.ro (corresponding author)}
\author{M. Cernaianu}
\affiliation{ELI-NP, ``Horia Hulubei" National Institute for Physics and Nuclear Engineering, 30 Reactorului Street, RO-077125, Bucharest-Magurele, Romania}

\author{D. Doria}
\affiliation{ELI-NP, ``Horia Hulubei" National Institute for Physics and Nuclear Engineering, 30 Reactorului Street, RO-077125, Bucharest-Magurele, Romania}
\author{P. Ghenuche}
\affiliation{ELI-NP, ``Horia Hulubei" National Institute for Physics and Nuclear Engineering, 30 Reactorului Street, RO-077125, Bucharest-Magurele, Romania}

\author{V. Horný}
\affiliation{ELI-NP, ``Horia Hulubei" National Institute for Physics and Nuclear Engineering, 30 Reactorului Street, RO-077125, Bucharest-Magurele, Romania}

\date{\today }

\begin{abstract}
 We propose a novel scheme for the population and depletion of nuclear isomers. The scheme  combines the 
 $\gamma$-photons with energies $\gtrsim \SI{10}{keV}$ emitted during the interaction of a contemporary high-intensity laser pulse with a plasma and one or multiple photon beams supplied by intense lasers. Due to nonlinear effects, two- or multi-photon absorption dominates over the conventional multi-step one-photon process for an optimized gamma flash. Moreover, this nonlinear effect can
be greatly enhanced with the help of externally supplied low-energy photons coming from another laser. These low-energy photons act such that the effective cross-section experienced by the $\gamma$-photons becomes tunable, growing with the intensity $I_0$ of the beam. 
Assuming $I_{0}\sim \SI{e18}{Wcm^{-2}}$ for the photon beam, an effective cross-section as large as \SI{e-21}{cm^2}
to \SI{e-28}{cm^2} for the $\gamma-$photon can be achieved. Thus, within state-of-the-art \SI{10}{PW} laser facilities,
the yields from two-photon absorption can reach \SI{e6}{} to \SI{e9}{} isomers per shot for
selected states that are separated from their ground state by E2 transitions.
Similar yields for transitions with higher multipolarities can be accommodated by multi-photon absorption with additional 
photons provided.
\end{abstract}

\maketitle

\vspace{10mm}

\section{Introduction}
Nuclear isomers are excited states of nuclei that have a longer half-life ($t_{1/2} \gtrsim \SI{1}{ns}$) than the states in the prompt decay paths of a nucleus following an induced excitation. 
Isomers with $t_{1/2}> 10$ years (e.g., $^{93m}$Nb, $^{113m}$Cd, $^{178m}$Hf, etc.) are ideal candadids for nuclear batteries that outperform conventional ones by $\times 10^6$ in energy density~\cite{Prelas2014}. Many isomers with $t_{1/2}\gtrsim$ hours have proven medical potentials~\cite{Nitipir2017,Klaassen2019,Sgouros2020}. Furthermore, isomers generally serve as pathways to enable nuclear lasing~\cite{RevModPhys.53.687,RevModPhys.69.1085}. 
Since their discovery~\cite{Hahn1921}, the efficient creation and induced depopulation of isomers has been one of the outstanding problems in physics. 
Any breakthrough on this topic will enable exciting applications in nuclear photonics. 

The production of isomers by brute force via photon absorption toward the desired state is inefficient and riddled with technological challenges. In general, it requires pumping through intermediate states with shorter lifetimes, except for a few cases where favorable transitions of ground-state to a higher excited state followed by its decay into the isomer state exist. Herein, the dilemma is that generating a large number of $\gamma$-photons to be absorbed resonantly by the nuclear level within a short time is difficult. Moreover, the direct pumping of a state exhibiting a longer half-life is challenged by the relatively narrow absorption bandwidth (e.g., \SI{10}{ps} correspond to a natural width of only $6.6\cdot \SI{e-5}{eV}$), which leads to the so-called graser dilemma, i.e., the half-life/width combinations given in nature require elusive laser pumping powers or the very unrealistic scenario of an induced nuclear explosion~\cite{RevModPhys.53.687,RevModPhys.69.1085}. As the energy gap between nuclear transitions often leads to isomers $\gtrsim \SI{100}{keV}$, no existing laser can be used in direct pumping, with a few exceptions such as, e.g., $^{229}$Th~\cite{Walker1999,PhysRevLett.132.182501}. 

Other than $\gamma$-excitation, alternative ways of generating isomers discussed in the past involve various nuclear reactions or Coulomb excitation, provided that 
suitable seed nuclei are available~\cite{PhysRevC.83.024604,Feng2022}. 
However, triggered depletion of isomers, ie, how to harvest energy with high efficiency, seems still an unsolvable problem~\cite{Chiara2018,Guo2021,PhysRevLett.128.242502,PhysRevLett.122.212501,Goldanskii1976,Morel2005,PhysRevLett.99.172502,PhysRevLett.83.5242,PhysRevLett.98.122701,PhysRevC.83.064617,Carroll2013,Kirischuk2015,vonderWense2016,Carroll2009,PhysRevE.97.063205,PhysRevLett.120.052504,PhysRevLett.127.042501,PhysRevC.58.333,PhysRevC.58.2560,PhysRevC.64.061302,PhysRevC.71.024606,Walker1999,Spohr:2023koj}. Nevertheless, proposals that utilizing intensive ($\gtrsim\SI{e10}{Wcm^{-2}}$) optical photons toward a virtual state followed by spontaneous emissions to the desired state provide an interesting alternative~\cite{PhysRevC.23.50,PhysRevC.20.1942,PhysRevC.23.1007,PhysRevC.29.1124}.

High-power laser systems (HPLS) can become core entities to spearhead associated developments~\cite{PhysRevLett.127.052501,PhysRevC.100.064610,PhysRevC.101.044304,von2020theory,PhysRevC.106.024606,PhysRevC.105.054001,PhysRevC.99.044610,PhysRevC.100.041601,Li2021,PhysRevC.104.044614,Lv2022,PhysRevA.108.L021502,PhysRevLett.131.202502} as they are now able to deliver intensities up to $\mathcal{P}\sim 10^{21-23}\SI{}{Wcm^{-2}}$~\cite{Yoon:19,Yoon:21,Li:18,Yu:18,ur2015eli,Tanaka2020} (see https://www.icuil.org/ for a comprehensive overview of current facilities).
When interacting with an overdense target, HPLS accelerate electrons in the materials to highly relativistic energies and drive strong currents in the plasma enviroment. The extreme, quasistatic magnetic fields then interact with energetic electrons, generate dense $\gamma$-photons~\cite{PhysRevLett.108.195001,PhysRevLett.108.165006,Ji2014,PhysRevLett.112.015001,PhysRevLett.115.204801,Chang2017,Zhu2018,Huang2019,PhysRevA.83.032106,PhysRevX.7.041003,Snyder2019,Shen2024} and rejuvenates the hope of a mass production of isomers. 
However, existing schemes of manipulating isomers incorporate single- or multi-step excitation via very dense $\gamma$-photons to some ``stepping" intermediate states in the decay chain toward the desired state~\cite{PhysRevC.58.333,PhysRevC.58.2560,PhysRevC.61.044303,Feng2023}, which, even with an optimal HPLS-based $\gamma$-production, give yield estimates that are far from even suggesting a technical realization \cite{Borghesi2006,roth2002energetic,logan2006assessment,Roth2008, PhysRevX.11.041002,Sarma2022,Snyder2019,Wang2023,EftekhariZadeh2022,Zhao2022}. Reaching a sufficient population yield for the isomer population remains an insurmountable challenge.     


\begin{figure}[t]
\includegraphics[width=0.47\textwidth,clip=true]{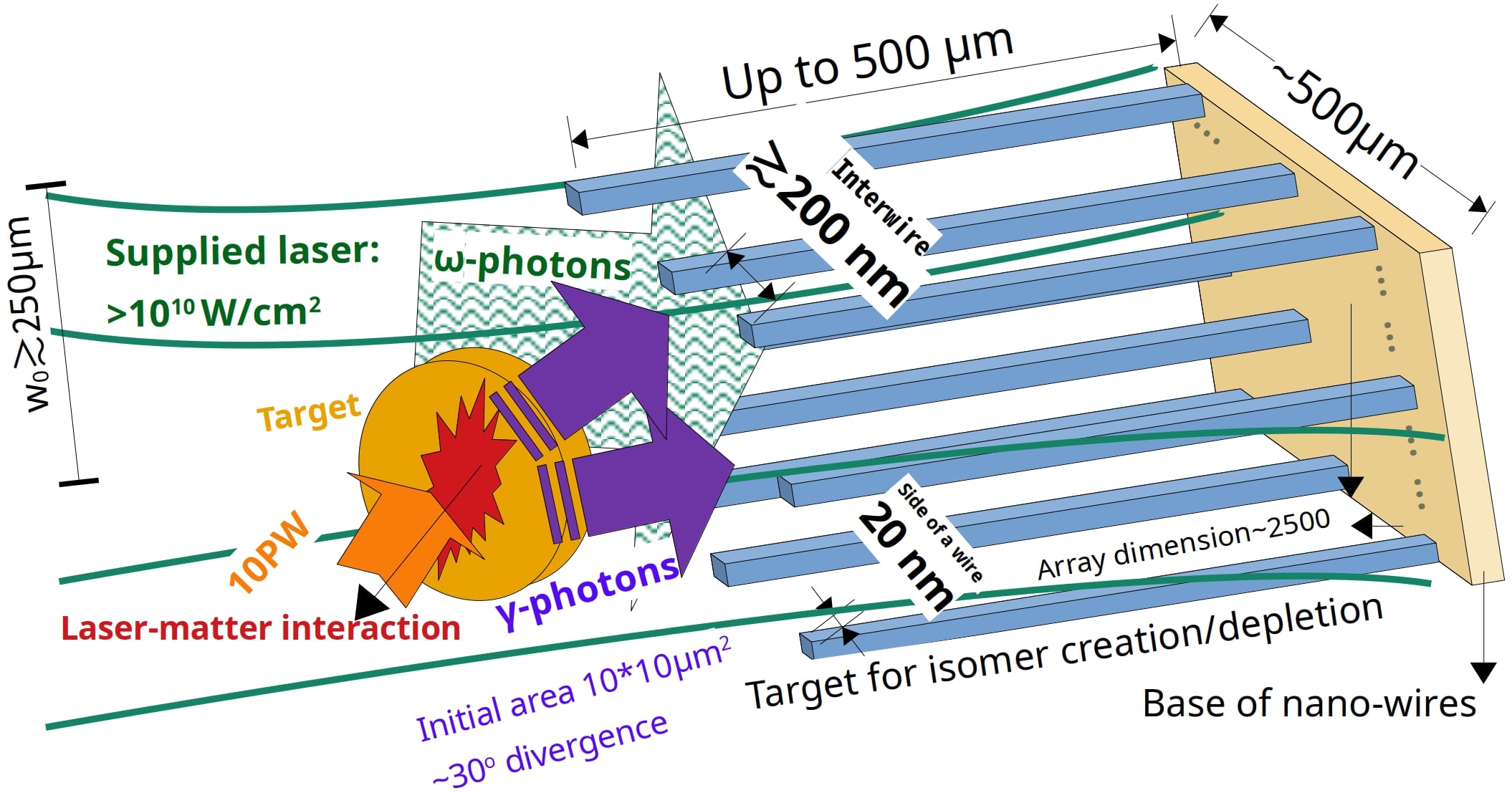}
\caption{Schematic illustration of the proposed scheme with a possible target design. Note that the two-directional $\gamma$-photons corresponds to the laser-matter interaction scheme which has the most intense $\gamma$-flux in Fig.~\ref{fig2}, while in general the $\gamma$-photons can have 
varied characteristics. (Dimensions not to scale, see supplemental materials for reasons behind the given parameters) 
}
\label{fig_tar}
\end{figure}
 
In this work, we demonstrate that the efficiency of isomer pumping and successive depletion can be drastically improved ($\ge \SI{e10}{}$-fold) by combining the $\gamma$-photons generated by an HPLS with intense supplied optical/infrared photons ($\equiv \omega$-photons), through a two- or multi-photon absorption (nPA) mechanism, as illustrated in Fig.~\ref{fig_tar}. This nPA mechanism prescribes a non-linear growth of yields~\cite{GM,PhysRevLett.7.229,PhysRevLett.42.1397,PhysRevC.20.1942,PhysRevC.23.50,PhysRevC.23.1007,Collins1982} and
becomes dominant as the intensity of the incident photons increases. 

\section{Theory}
Denoting $E_i$ and $I_i$ the energy and the corresponding intensity (note that $I_i$ is the number of photons per area per time, while $\mathcal{P}_i\equiv I_i E_i$ is the energy flux---which is later denoted as intensity in the unit of \SI{}{Wcm^{-2}}) of the $i^{th}$ photon category in an nPA process. Assuming the energies are tunable so that $E\equiv\sum_{i=1}^{n} E_i$ equals the gap between the initial state and the state one wishes to populate,
then 
\begin{equation}
\frac{d\mathcal{Y}_{(n)}}{dz}=N\cdot(I_1I_2\cdot...\cdot I_n\sigma_{npa}+...),\label{eq1}
\end{equation}
where $\mathcal{Y}_{(n)}$ denotes the yields of the final state per unit time given by $n$-photon absorption, 
$N$ is the number of nuclei photons encountered per distance in the target $dz$, and 
$\sigma_{npa}$ is the generalized cross-section for nPA and is a function of each $E_i$. 
The unit of 
$\sigma_{npa}$ is [length]$^{2n}$[time]$^{n-1}$, i.e.,
a direct extension to the well-known 2PA cross-section
($\sigma_{2pa}=\sigma_{GM}$)\cite{GM}. 
Note that nPA corresponds to iterations on the interaction Hamiltonian $H_{int}$---which is normally 
further expanded into electromagnetic multipoles. Therefore, unless the product of intensities reaches a sufficient
threshold, $I_1I_2\cdot...\cdot I_n\sigma_{npa}$ is negligible compared to 1PA with an $H_{int}$ equal to the multipolarities of the transition. 
Conversely, additional terms (denoted as ``..." in Eq.~(\ref{eq1})) will dominate over $I_1I_2\cdot...\cdot I_n\sigma_{npa}$ if any of the $I_i$ satisfies
\begin{equation}
\sigma_{npa} < I_i\sigma_{(n+1)pa}. \label{eq2} 
\end{equation}
In this case, two photons (instead of one) in the $i^{th}$ category being absorbed at once will be preferred, and the
original nPA process is surpassed by a (n+1)PA process.
The fundamental reason for this enhancement lies in combinatorial enhancement as illustreated in Fig.~1 of Ref.~\cite{yang2025nuclearphysicslowenergyhigh}, which makes multiphoton absorption more efficient than single-photon absorption under an intense environment. 
One can rewrite $I_1I_2\sigma_{2pa}$ as $I_1\sigma_{eff}(I_2)$ for 2PA. Thus,
the effective cross-section $\sigma_{eff}(I_2) $ experienced by those $I_1$-photons becomes tunable and depends on $I_2$. The dependence is linear only when $I_2$ is low, i.e., $\sigma_{eff}(I_2)\sim I_2\sigma_{2pa}$, but becomes $\sim I_2^2\sigma_{3pa}$ when $I_2$ reaches a critical value so that $\sigma_{2pa} < I_2\sigma_{3pa}$.
For convenience, we assume that all of the $I_i$ in Eq.~(\ref{eq1}) are below their critical values governed by Eq.~(\ref{eq2}). Any exceeding $I_i$ will be re-distributed in favor of a new category (n+1), as (n+1)PA dominates in this case. In that case, one can define the effective cross-section experienced by $I_1$ (via nPA) as
\begin{eqnarray}
\sigma^{npa}_{eff}(I_2,I_3,...,I_n)= I_2\cdot...\cdot I_n\sigma_{npa}, \label{eq3} 
\end{eqnarray}
where the unit of $\sigma^{npa}_{eff}$ is [length$^2$].
The remaining task is to estimate the size of $\sigma_{npa}$. The concept of nPA has been widely practised in atomic and molecular physics, though it has yet to be observed experimentally in nuclear systems. Only reverse processes have been observed based on double $\gamma$-decay~\cite{PhysRevLett.53.1897,Kramp1987,PhysRevLett.7.170,Alburger1964,Harihar1970,Nakayama1973,Vanderleeden:1970zn,PhysRevLett.53.1897,Soderstrom:2020iaz,freirefernández2023measurement}. Nevertheless, the theoretical derivation first introduced by G\"{o}ppert-Mayer~\cite{GM} and its later extensions~\cite{LAMBROPOULOS197687,Friedrich2017,Delone2000} are very general and do not rely on whether the systems at
hand are nuclei or atoms. Denoting $e$ as the charge, $|i\rangle$, $|m\rangle$ and $|f\rangle$ as the initial, intermediate, and final state in the process, the transition rate of 2PA reads~\cite{supplemental},
\begin{equation}
R_{2pa}=\frac{e^4 \mathcal{E}^2_1 \mathcal{E}^2_2}{16\hbar^4}|\mathcal{M}|^2 2\pi\delta_t(\omega-\omega_1-\omega_2), \label{prob}
\end{equation}
where $\omega_i=E_i/\hbar$, with $E_i$ the energy of the
$i^{th}$-category photons with amplitude $\mathcal{E}_i$, $\hbar$
the reduced Planck constant. $\delta_t$ has the dimension of [time], and
\begin{equation}
\mathcal{M}=\sum_m\left[\frac{\langle f|\hat{H}_{2}|m\rangle\langle m|\hat{H}_{1}|i\rangle}{\omega_1-\omega_{mi}}+(1\longleftrightarrow2)\right], \label{m}
\end{equation}
where $\omega_{mi}=\omega_m-\omega_i$ is the energy gap between the initial and the $m$ state, with $m$ representing a physical state which belongs to any eigenstate of the nuclear Hamiltonian (including $i$ or $f$ itself). To maximize the transition amplitude $|\mathcal{M}|$, the virtual state after each photon absorption needs to resonate with a physical state to minimize the denominators of Eq.~(\ref{m}).

In the atomic case, the wavefunctions of excited-states can be calculated accurately~\cite{johnson1980relativistic,slater1951simplification,jonsson2019computational,wilson1985diagrammatic,manson2009relativistic}, which enables an evaluation of $|\mathcal{M}|$~\cite{PhysRevA.22.1576,PhysRevA.30.2805,PhysRevA.34.199,PhysRevA.34.185,Mainfray1991,Friedrich2017,Delone2000,collins1997progress,PhysRevB.104.085201}. 
Unfortunately, most excited states of nuclei cannot be accurately described due to a combination of a complicated structure of nuclear forces and the difficulty of solving the nuclear many-body problem~\cite{vanKolck:2020llt,Yang:2019hkn,Yang:2021vxa,Tews:2022yfb}. In this work, we do not focus on specific nuclei and adopt the Weisskopf estimates~\cite{Blatt:1952ije} to evaluate the transition operators $\langle\hat{H}_{i}\rangle$ due to the $i^{th}$ photon. For convenience, we define $i=1$ as the
$\gamma$-photons and $i=2$ as the $\omega$-photons.
 
We consider two cases for 2PA. Case (a) represents the situation that no intermediate state within $\sim1$ keV from $\langle i|$ or $\langle f|$ exists. Therefore, the virtual state needs to resonate with either $\langle i|$ or $\langle f|$ itself, causing the
denominator in Eq.~(\ref{m}) to become $\omega_2$ (i.e., the one labeled as the supplied photon). The corresponding transition $\hat{H}_{2}$ must be at least M1 or higher, as $\langle m|$ is represented by either $\langle i|$ or $\langle f|$ and the angular momentum quantum number between the transition $\Delta J^{\pi}=0^+$. Furthermore, one needs to exclude the transition from $J=0$ to $J=0$, which is forbidden for a single photon in the first order~\cite{PhysRevLett.53.1897}. For $\hat{H}_{1}$, E1 gives the largest possible matrix element. 
Case (b) has $\langle m|$ separated from $\langle i|$ or $\langle f|$ by $\Delta J^{\pi} = 1^{-}$ and $\Delta E\lesssim 1$ keV. In this case, each photon can undergo E1 transition. 

The difference between case (a) and (b) is whether the $\omega$-photons perform M1 or E1 transitions. As the typical ratio of transition probabilities between E1 and M1 ranges from $10^1-10^3$, case (b) will become negligible compared to (a) if the aforementioned gap $\Delta E> 1$ keV, due to a larger denominator in Eq.~(\ref{m}).



We note that $\langle i|$ and $\langle f|$ can be any eigenstate of the nuclear Hamiltonian.
For isomer creation, $\langle f|$ is the isomer or a higher excited state that has a favorable decay path to the desired isomer, and $\langle i |$ is the initial state. For depletion, $\langle i|$ is the isomer and $\langle f|$ is the state one wishes to arrive, which can be any excited state with $J^{\pi}$ favorable for decaying into the ground-state. 

\section{Yields estimate}

The realistic cross-section of 2PA is conventionally expressed as~\cite{PhysRevA.34.199}
\begin{eqnarray}
\sigma_{2pa}=\sigma_{0}gG, \label{cross}
\end{eqnarray}
with $\sigma_{0}$ (unit: [length]$^4$) the line-shape-independent cross-section, $g$ (unit: [time]) the line-shape function, and $G$ the photon statistical factor~\cite{PhysRevA.1.1445,Zoller1980}. $G=1$ for a single-mode laser and becomes larger in multi-mode cases. The value of $g$ can be estimated by a Gaussian distribution with a full width at half maximum
(FWHM) equal to the greater of the width ($\equiv w_>$) of $\langle f|$ or the laser bandwidth.
\begin{eqnarray}
g=2\sqrt{\frac{\ln2}{\pi}}\frac{1}{w_>}\sim \frac{0.939}{w_>}. \label{gw}
\end{eqnarray}
One can see that $\sigma_{2pa}$ is weakened by $w_>$, which is dominated by recoil broadening and should not exceed \SI{1}{eV} for solid targets. Using ultra-intense beams could potentially broaden $w_>$~\cite{Hack1961} and free the nuclei in the solid target for an intensity $\gtrsim \SI{e18}{Wcm^{-2}}$. 

With detailed derivations given in the supplemental materials~\cite{supplemental},
the effective cross-section(given in [cm$^2$]) experienced by the $\gamma$-photon reads
\begin{align}
\sigma^{2pa,E1+E1}_{eff}&\approx 3\cdot 10^{-51}\frac{E_{\gamma}}{w_>}\mathcal{P}_2 \frac{A^{4/3}}{(\Delta E)^2}G , \label{eff_e1} \\
\sigma^{2pa,M1+E1}_{eff}&\approx 5\cdot 10^{-51}\frac{E_{\gamma}}{w_>}\mathcal{P}_2 \frac{A^{2/3}}{(E_2)^2}G ,
\label{eff_m1}
\end{align}
where the superscripts $M1+E1$ and $E1+E1$ represent cases (a) and (b), respectively. $E_{\gamma}$ and $w_>$ must have the same unit, $\mathcal{P}_2$ is the intensity of the supplied laser in \SI{}{Wcm^{-2}}, $\Delta E=|E_m-E_{i/f}|$ (in eV) is the gap between $\langle m|$ and the nearest of the initial/final state, which reduces to $E_2$ in the case of $M1+E1$. $A$ is the number of nucleons.

Eq.~(\ref{eff_e1}) and (\ref{eff_m1}) agree with the estimates in Refs.~\cite{PhysRevLett.42.1397,PhysRevC.20.1942} if one adopts $G=1$ and the natural width of $\langle f|$ for $w_>$. However, this is likely to apply only in the cases of M\"{o}{\ss}bauer nuclei~\cite{Mssbauer1958}. In general, $w_>$ is likely to be $10^{3\sim 4}$ times larger. Thus, the effective cross-section (\SI{e-26}{cm^2} to \SI{e-32}{cm^2}  for $A\approx200$ nuclei) given before by assuming $\mathcal{P}_2=\SI{e10}{Wcm^{-2}}$ will be suppressed accordingly. However, today's lasers can reach $\sim \SI{e23}{Wcm^{-2}}$. Therefore one can easily compensate for the  $\sim 10^{3}$ to $10^4$ suppression-factor and obtain $\sigma^{2pa}_{eff}\approx\SI{e-21}{cm^2}$ to \SI{e-28}{cm^2} with $\mathcal{P}_2\lesssim \SI{e18}{Wcm^{-2}}$. A higher $\mathcal{P}_2$, though achievable today, complicates the yield estimate, as target disintegration needs to be considered. Nevertheless, this process (mechanical shock-wave propagation) is much slower than the speed of light and the nuclear transition time, so pumping/depletion will still occur. 
\begin{figure}[t]
\includegraphics[width=0.47\textwidth,clip=true]{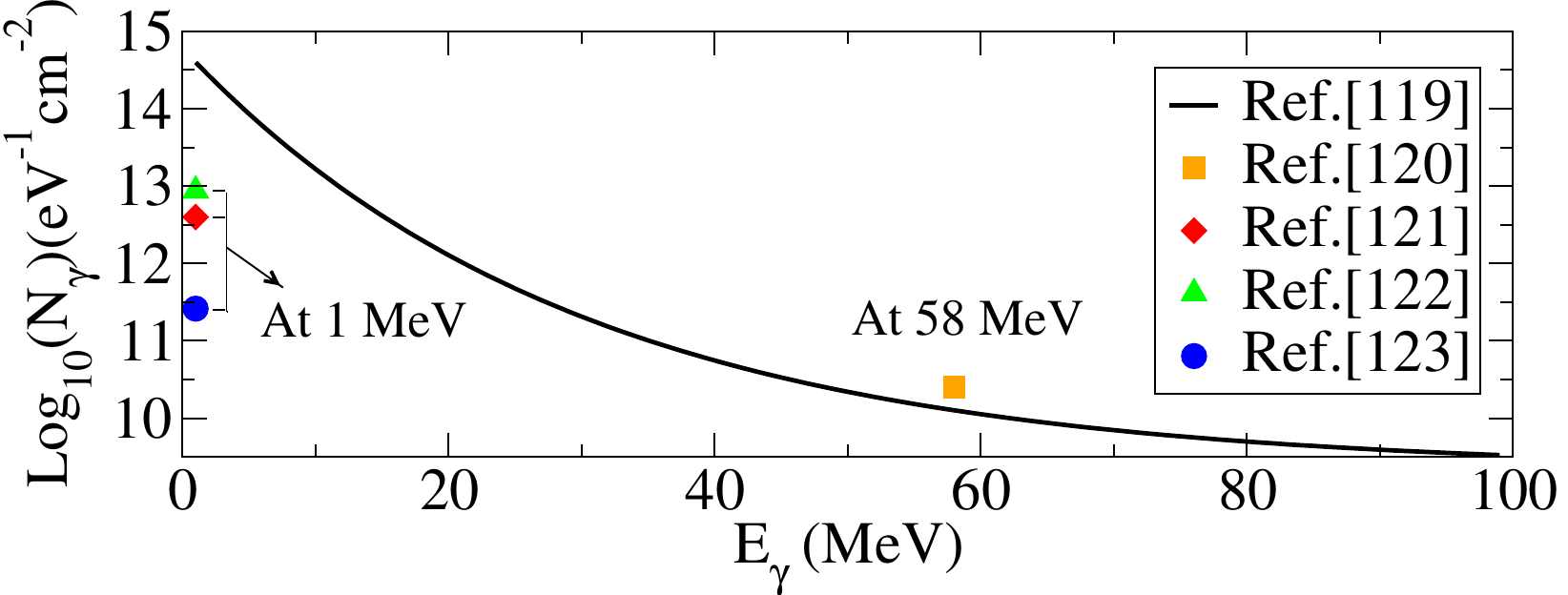}
\caption{Number of $\gamma$-photons $N_{\gamma}$ per \SI{}{eV} \SI{}{cm^2} as a function of energy generated via laser-plasma interaction in a single shot converted from Refs.~\cite{PhysRevApplied.13.054024,Gu2018,Xue2020,PhysRevE.104.045206,Heppe2022}, with details listed in supplemental material~\cite{supplemental}. Note that the actual $\gamma$-flux occupies an area of only $\approx \SI{100}{\upmu m^2}$. The accumulated time is $t\approx$ \SI{15}-\SI{50}{fs}.}
\label{fig2}
\end{figure}

Note that the conventional treatment of $gG\approx1/w_>$ in Eq.~(\ref{cross}) reduces the effective cross-section due to the assumption that the beams are near \textit{monochromatic}. This is not the case for the $\gamma-$source generated via laser-plasma interaction, as shown in Fig.~\ref{fig2}. Unlike conventional beams, the spectrum is continuously distributed from \SI{1}{keV} to \SI{50}{MeV} \cite{PhysRevApplied.13.054024,Gu2018,Xue2020,PhysRevE.104.045206,Heppe2022,Jirka2020}. Yield estimates, in this case, require special care. The line shape of the final state does not play the same role as before, as each shifted or broaden-state will always have a combination of $\gamma-$photon plus $\omega$-photon centered to its peak. 
A more convenient treatment is considering the integrated intensity of $\gamma$-photons over the energy interval equal to the total width $w_>$ of the final state. Defining such integrated intensity as $\Delta I_{\gamma}$, as all $\gamma$-photons within that energy interval can excite the nuclei, the yield as defined in Eq.~(\ref{eq1}) becomes
\begin{eqnarray}
\frac{d\mathcal{Y}_{(2)}}{dz}=N \Delta I_{\gamma}\frac{w_>}{\bar{w}} \sigma^{2pa}_{eff}, \label{yie}
\end{eqnarray}
with $\bar{w}$ the bandwidth of the supplied laser. 

\subsection{Yields soley from a single $\gamma$-flash}
We first investigate whether 2PA already dominates over 1PA under a single $\gamma$-flash from the laser-matter interaction.
With the $\gamma$-flash alone, photons within the narrow energy range of $E_{\gamma}\in[1,1.001]$ keV reach an area density of $N_{\gamma}\sim10^{12}$ cm$^{-2}$ even estimated very conservatively from Fig.~\ref{fig2}, which corresponds to $\mathcal{P_{\gamma}}\approx $\SI{e9}{Wcm^{-2}}. Plugging $\mathcal{P}_{\gamma}$ as $\mathcal{P}_2$ into Eq.~(\ref{eff_e1}), the effective cross-section for an incoming \SI{1}{MeV} $\gamma$-photon assisted by \SI{1}{keV} photons becomes $\sigma^{2pa}_{eff}\approx 10^{-37}$ cm$^2$. Here, $\sigma^{2pa}_{eff}$ is further suppressed than previously estimated as $\Delta E$ or $E_2\sim$\SI{1}{keV}. This leads to a yield $\sim3\cdot10^{-6}$ per cm in the target just by considering the combination of $E_{\gamma1}=\SI{1}{keV} $ and $E_{\gamma2}=\SI{1}{MeV}$. Assuming similar $N_{\gamma}$ persists from \SI{1}{keV} to \SI{1}{MeV}, then $10^6/2$ combinations are available. After summing each combination and considering the energy dependence, $10^{-4}$ isomers per shot are generated within \SI{1}{cm} thickness in the target via 2PA. While the yields under the same condition from sequential 1PA+1PA process is $10^{-9}$ (assuming the cross-section at each step is 1 barn, a typical width ($\approx10^{-4}$ eV) for the intermediate state, a negligible angular-divergence of $\gamma$, and an absence of favorable decay/feeding paths from higher-states).
Thus, 2PA dominates over sequential photon absorption, considering only the $\gamma-$flash.
In summary, the factors contributing to the difference in this specific case are:
\begin{itemize}
\item The width of the intermediate state 1PA+1PA passes through, which gives a $10^{-4}$ factor
in its final yield with respect to the value without reduction.
\item The continuous energy spectrum of the $\gamma$-flash, which enables various combinations of the
photon energies within 2PA. This gives another $\approx10^2$ enhancement in the yield from 2PA.
\end{itemize}

\subsection{Yields from a combination of $\gamma$-flash and low-energy photons supplied by lasers}
\begin{figure}[t]
\includegraphics[width=0.47\textwidth,clip=true]{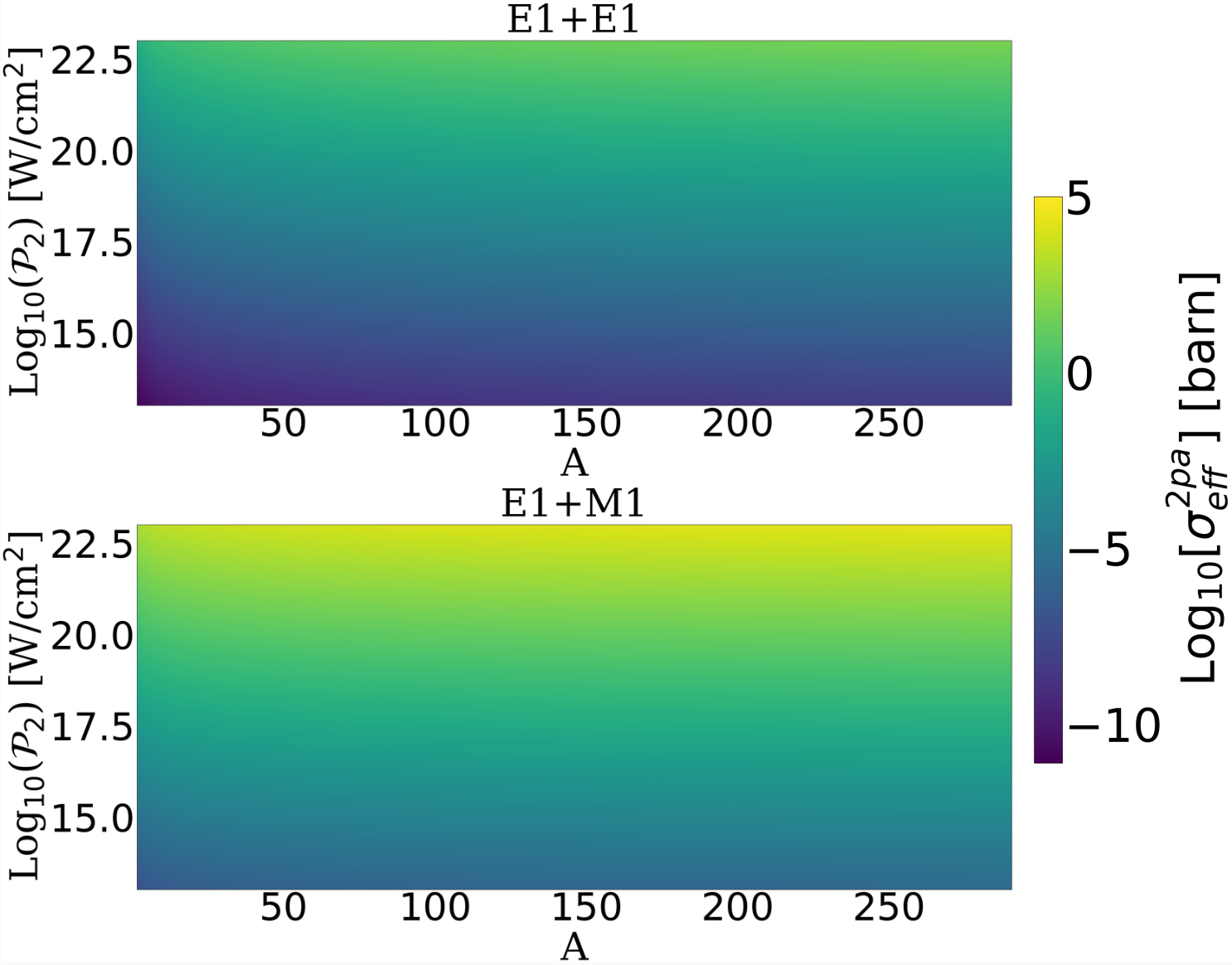}
\caption{The 2PA effective cross section as a function of mass number $A$ and the intensity of the supplied laser $\mathcal{P}_2$, where the upper and lower panels correspond to Eq.~(\ref{eff_e1}) with $\Delta E=100$ eV and Eq.~(\ref{eff_m1}) with $E_2=1$ eV, respectively. Other parameters are $E_{\gamma}=1$ MeV, $\omega_>=1$ eV, $G=1$.}
\label{sigmaplot}
\end{figure}

Combining the $\gamma-$photon with $\omega$-photons supplied at $\mathcal{P}\gtrsim \SI{e10}{Wcm^{-2}}$ further increases the yields, as a $\gtrsim \SI{e3}{}$ enhancement coming from the reduction of bandwidth from the $\gamma-$spectrum to the supplied laser, and another up to \SI{e6}{} (system-dependent) enhancement due to a smaller $E_2^2$ or $(\Delta E)^2$. The effective cross section currently achievable as a function of $A$ and $\mathcal{P}_2$ is plotted in Fig.~\ref{sigmaplot}. With the setup illustrated in Fig.~\ref{fig_tar}, with two modern HPLS, the number of isomers could easily reach \SI{e6}{} per shot ($\lesssim \SI{50}{fs}$). 
In fact, we are mainly limited by the total number of $\gamma$-photons that are actually generated per absorption width, not the effective cross-section, which is sufficiently large starting from $\mathcal{P}\approx   \SI{e13}{Wcm^{-2}}$ so that most of the $\gamma$'s matching the excitation energy are absorbed within \SI{1}{cm} thickness in the target. Moreover, in the case of 2PA, we do not need to consider the reduction caused by the angular divergence of $\gamma$-photons as long as the whole propagation/spread stays within the target and is within the waist of the supplied laser, which is a highly desirable feature in contrast to sequential pumping. 

\section{Generalization}
\begin{figure}[t]
\includegraphics[width=0.4\textwidth,clip=true]{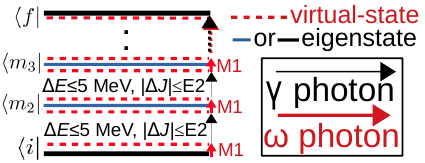}
\caption{Illustration of an nPA scheme, where the pumping is achieved through each virtual state with gaps equal to $E_{\gamma_i}$ (preferably $\lesssim5$ MeV) or $E_{\omega_i}$. The ``:" between $\langle m_3|$ and $\langle f|$ corresponds to ``..." in Eq.~(\ref{gen}) and denotes similar pumping processes utilizing the $\gamma_i+\omega_i$ pairs. Only virtual states located around $m_i$ contribute significantly.}
\label{fig_npa}
\end{figure}

 Now, we generalize the above ingredients to enhance more transitions. First, the magnitude of the achievable $\sigma^{2pa}_{eff}$ makes it possible to consider a $\langle \hat{H}_i \rangle$ of E2 type per step. A realistic E2 amplitude often exceeds the Weisskopf estimate and is suppressed only by $\sim \SI{e2}{}$ relative to E1 transitions due to collective excitation~\cite{Casten2001}. Thus, $\sigma^{2pa,E1+E2}_{eff}$ and $\sigma^{2pa,M1+E2}_{eff}$ would reach $\SI{e-23}{cm^2} \sim \SI{e-30}{cm^2}$ with $\mathcal{P}_2$ up to \SI{e18}{Wcm^{-2}}, while M2 and higher multipolarities are further suppressed by at least \SI{e6}{} in their magnitude and are unfavorable. Second, it is straightforward to generalize 2PA to nPA, i.e., just replacing Eq.~(\ref{prob}) by 
 \begin{equation}
R_{npa}=\frac{e^{2n} \mathcal{E}^2_1 \cdots \mathcal{E}^2_n}{4^n\hbar^{2n}}|\mathcal{M}^{(n)}|^2 2\pi\delta_t(\omega-\omega_1\cdots-\omega_n), \label{prn}
\end{equation}
with $\mathcal{M}^{(n)}$ as the sum over eigenstates $\langle m_{2}|\sim\langle m_n|$, i.e.,
\begin{widetext}
\begin{equation}
\mathcal{M}^{(n)}=\sum_{m_2}\cdots\sum_{m_n}\left[\frac{ \langle f|\hat{H}_{n}|m_n\rangle \cdots \langle m_3|\hat{H}_{2}|m_2\rangle\langle m_2|\hat{H}_{1}|i\rangle}{(\omega_1-\omega_{m_{2}i})(\omega_2-\omega_{m_3 m_2})\cdots(\omega_{n-1}-\omega_{m_{n}m_{n-1}})}+(\text{all permutation})\right]. \label{mn}
\end{equation}
\end{widetext}
Note that, like in Eq.~(\ref{m}), the adopted-intermediate state $\langle m_i|$ can be the same as $\langle i|$ or $\langle f|$. Maximum resonance occurs when each virtual state is separated from an intermediate state by the natural width of the intermediate state.  
A favorable scenario is that
each $\gamma_i$-photon (performs E1 or E2 transition) is assisted by one $\omega_{i}$-photon (performs M1 transition) to reach a virtual state, i.e.,
\begin{equation}
\underbrace{\gamma_1+\omega_1}_{|\Delta J|\leq E2}+\underbrace{\gamma_2+\omega_2}_{|\Delta J|\leq E2}+... ,\label{gen}
\end{equation}
where $|\Delta J|\leq E2$ means the transition is within $|\Delta J|\in$(E1,M1,E2).
As illustrated in Fig.~\ref{fig_npa}, in general, the largest contribution facilated by nPA would be that all of the $\omega$-photons are used in resonating with each intermediate state via M1 transitions. Each leap with $1$ keV$\lesssim \Delta E\lesssim 10$ MeV is completed by the $\gamma$-photons from the $\gamma$-flash, which connects intermediate states $\langle m_i|$ separated by E1, M1 or E2 transitions. The total number of intermediate states as described in Fig.~\ref{fig_npa} is $n/2-1$ for nPA, with the net angular momentum difference between $\langle i|$ and $\langle f|$ up to $|\Delta J|=n$. 
Possible modifications to the above scheme include the scenario of the existence of adjacent intermediate states separated by an energy comparable to the energy of $\omega$-photons and with $|\Delta J^{\pi}|=1^-$, which then favors an E1 transition for the $\omega$-photons.

With the general nPA scheme in hand, we now estimate the maximum value of n that can be achieved today.
Note that the factor which governs the strength between nPA and 2PA is~\cite{graserour}
\begin{equation}
    3\cdot 10^{-24}\mathcal{P}_\mathrm{i} \,\frac{X_\mathrm{i}}{(E_{\omega})^2},
    \label{npa_a}
\end{equation}
where the subscript $i$ runs from 1 to $n-1$ for nPA, $X_\mathrm{i}=A^{2/3}$ (with $A$ the total number of nucleons) if the transition carried by the $i^\mathrm{th}$-photon is of E1 type. For M1 and E2 type transitions, $X_\mathrm{i}$ will be suppressed by $\gtrsim25$ and $\gtrsim100$ times concerning the E1 value. $E_{\omega}$ (in eV) represents the energy difference between each virtual state and the state it resonates with. 
The condtion for nPA$\approx$(n-1)PA corresponds to Eq.~(\ref{npa_a})$\approx1$. For $A\approx200$, the scenario described by Fig.~\ref{fig_npa} then requires $\mathcal{P}_i\approx 10^{23}$ W/cm$^2$ (for both $\gamma$- and $\omega$-photons) if $ E_{\omega}=1$ eV, or $\mathcal{P}_i\approx 10^{17}$ W/cm$^2$ if $ E_{\omega}=10^{-3}$ eV. Note that $E_{\omega}=10^{-3}$ eV corresponds to a value slightly greater than the typical width of the nuclear intermediate states. To achieve such a pumping, the supplied $\omega$-photons need to come from an intense far-infrared/terahertz source, so that the gaps between virtual states in Fig.~\ref{fig_npa} can be of the order $10^{-3}$ eV. A further reduction of the gaps $\lesssim10^{-4}$ eV would start to push the virtual states to overlap with the physical states and terminate the nPA process in the intermediate state---which can be perceived with the scenario that the two nearby red dashed-lines overlap with the blue solid-line in Fig.~\ref{fig_npa}.

As shown in Fig.~\ref{fig2}, the power density of the $\gamma$-photons---though not reaching the critical value for nPA$\approx$(n-1)PA---is rather intense, i.e., $\mathcal{P_{\gamma}}\sim \SI{e9}{Wcm^{-2}}$ to $\SI{e12}{Wcm^{-2}}$ for E$_{\gamma}\leq \SI{1}{MeV}$.
Thus, combining a typical $\gamma$-flash and $\omega$-photons coming from another laser, 4PA can be achieved today in $A\approx200$ nuclei with an effective cross section $\approx10^{-25}$ cm$^2$. 
Typically, an isomer separated by $|\Delta J|\gtrsim4$ concerning all states to which it can decay could already have a half-life $\gtrsim \SI{1}{year}$, which would be accessible with our proposed population scheme in a site equipped with HPLS of PW class. Note that for isomer pumping, the $J^{\pi}$ structure of Fig.~\ref{fig_npa} suggests that $\langle f|$ would decay easily, so $\langle f|$ is mostly a state with favorable decay paths toward the isomer state instead of the isomer itself. We note that replacing the last step in Fig.~\ref{fig_npa} by an anti-Stokes process to reach the desired isomer state is possible and can be crucial for nuclear lasers, which we dedicated to a following work~\cite{graserour}. This feeding (through $\langle f|$) scheme also allows for an isomer population inversion against $\langle i|$. Counter-effects such as stimulated emissions or anti-Stokes processes
are suppressed before the population of $\langle f|$ dominates.

Experimentally, only two photon sources are required: an ultra-short \SI{10}{PW} laser pulse to generate $\gamma-$photons via laser-matter interactions, the other (with $\mathcal{P}\gtrsim \SI{e10}{Wcm^{-2}}$) supplies $\omega$-photons.
Two laser sources are sufficient because each of the $\omega$-photons will automatically select and combine with suitable $\gamma_i$-photons presented in Fig.~\ref{fig2} to achieve nPA. Thus, the $\omega_i$ can be the same in Eq.~(\ref{gen}) and be supplied by one source.
Using more than one $\omega-$source is beneficial provided that it can be tuned to another energy (away from the $1^{st}$ one by at least $w_>$), then the $\gamma$-photons in another energy interval can be utilized to increase the yields.

\section{Experimental realization}
Our scheme applies to gas, liquid, and solid isomer targets. Depending on the shape/thickness of the target used to generate $\gamma$-photons, a filter which acts as a plasma mirror might be necessary to prevent the residue PW-pulses (after laser-matter interaction) from destroying a solid isomer-target. For solid isomer-targets, two scenarios can occur: (1) The target remains transparent for the $\omega$-photons---which typically corresponds to non-metallic targets when $\mathcal{P}\lesssim \SI{e11} - \SI{e12} {Wcm^{-2}}$. In this case, all photons penetrate the target, and no special design is required. (2) For cases where the effect of free-electrons dominates, which corresponds to metallic targets or any highly ionized target when $\mathcal{P}\gtrsim \SI{e12}{Wcm^{-2}}$. In this case, the $\omega$-photons can penetrate only within a skin depth, which is usually a few nm. To maximize the yields, we need special target designs. One possibility, as sketched in Fig.~\ref{fig_tar}, is a nano-wire-like structure, which allows the laser to propagate along the surface of the wires for a longer distance, and therefore maximize the yields~\cite{Purvis2013,Bargsten2017,Cao2010,Samsonova2018,Kulcsr2000,Cristoforetti2014,Hollinger2017,Ivanov2017,Khaghani2017,Gizzi2020,Cristoforetti2020,Sarkar2017,Moreau2019,Hollinger2020,Park2021,Pan2024,Xia2024,Yang2023,Kong2022,Chao2022,EftekhariZadeh2022}. The realistic waist of the $\gamma$-photons is $\sim \SI{5}{\upmu m}$. 
One possibility to optimize the yield
will be an array of square-type nano-wires, each with dimension $\sim \SI{20}{nm}$ on each side, and with a sufficient inter-wire distance ($\gtrsim200$ nm) 
to allow free propagation of photons.        

\section{Summary}
Note that the fascinating idea of enhancing cross-sections by supplying various intensive beams is not new~\cite{Arad1979,BECKER1984441,PhysRevLett.42.1397,PhysRevC.20.1942,PhysRevC.23.50,PhysRevC.23.1007,PhysRevC.29.1124,PhysRevA.99.042517,ishkhanov1980excitation,Baklanov1976,Winterberg1986,eliezer1995induced,eliezer1995possibility,Rivlin2004}. However, we demonstrate here for the first time that the feasibility of such a mechanism can be carried out with the help of the $\gamma-$photons generated via laser-matter interaction. Therefore, we do not need to rely on any spontaneous emission coming from the anti-Stokes or Raman process, as in those previous depletion proposals. 

Moreover, previous methods utilizing sequential pumping suffer from various restrictions from
the intermediate states and can be very inefficient unless there exist favorable transitions of the ground state to a higher excited state followed by its decay into the
isomer state.
In contrast, our scheme can be applied to pump and deplete a very wide class of nuclear isomers, and therefore represent a crucial step toward realizing exciting new concepts in nuclear photonics.

\vspace{0.5cm}
\begin{acknowledgments}
We thank Paolo Tomassini and Bogdan Corobean for useful discussions and suggestions. 
This work was supported by the Extreme Light Infrastructure Nuclear Physics (ELI-NP) Phase II, a project co-financed by the Romanian Government and the European Union through the European Regional Development Fund - the Competitiveness Operational Programme (1/07.07.2016, COP, ID 1334);  the Romanian Ministry of Research and Innovation: PN23210105 (Phase 2, the Program Nucleu), the ELI-RO grant Proiectul ELI-RO/RDI\_2024\_AMAP, ELI-RO\_RDI\_2024\_LaLuThe, ELI-RO\_RDI\_2024\_SPARC and ELI10/01.10.2020 of the Romanian Government; the European Union, the Romanian Government and the Health Program, within the project ``Medical applications of high-power lasers - Dr. LASER"; SMIS Code: 326475 and the IOSIN funds for research infrastructures of national interest.
We acknowledge EuroHPC Joint Undertaking for awarding us access to Karolina at IT4Innovations (VŠB-TU), Czechia, under project number OPEN-34-63 and EHPC-REG-2023R02-006 (DD-23-157), and CINECA HPC access through PRACE-ICEI standard call 2022 (P.I. Paolo Tomassini).
\end{acknowledgments}


\subsection*{Conflict of Interest}
The authors have no conflicts to disclose.




\section*{Supplemental material}

\section{Derivations of two-photon transition rate}

The system is perturbed by an electric field given by lasers or $\gamma$-sources consists of photons, i.e., 
\begin{eqnarray}
H_{int}=\sum_{i=1}^{n}\mathcal{E}_i\hat{e_i}\cos[\vec{k_i}\vec{r}-\omega_i t]. \label{eq4}
\end{eqnarray}
Here $|\vec{k_i}|=E_i/\hbar c$, $\omega_i=E_i/\hbar$, with $E_i$ the energy of photons in the i$^{th}$-category with amplitude $\mathcal{E}_i$ propagates in direction $\hat{e_i}$, $\hbar$ the reduced Planck constant and c the speed of light. $t$ is the time. In this work we will focus on the absorption case, thus keeping only the $e^{-i\omega_i t}$ component of the cosine function in Eq.~(\ref{eq4}), and we consider each single-photon transition through intermediate states is of dipole (E1 or M1) form (i.e., expanding $e^{i\vec{k_i}\vec{r}}=1+i\vec{k_i}\vec{r}+...$ in Eq.~(\ref{eq4}) and keeping the components up to $O(\vec{k_i}\vec{r})$). Taking out the $\vec{k_i}\vec{r}$ part 
and denoting $e$ as the charge, $|i\rangle$, $|m\rangle$ and $|f\rangle$ the initial, intermediate, and final state in the transition process, the transition amplitude for 2PA can then be expressed as
\begin{widetext}
\begin{equation}
a^{[2]}(t)=\frac{e^2 \mathcal{E}_1 \mathcal{E}_2}{4\hbar^2}\sum_m \left[\frac{\langle f|\hat{H}_{2}|m\rangle\langle m|\hat{H}_{1}|i\rangle}{\omega_1-\omega_{mi}}\frac{e^{i(\omega-\omega_1-\omega_2)t}-1}{\omega-\omega_1-\omega_2}+(1\longleftrightarrow2)\right], \label{eq6}
\end{equation}
\end{widetext}
where the dipole interaction Hamiltonian reads
\begin{align*} 
\hat{H}_{i}&=\hat{e_i}\cdot \vec{r},   &\text{       
  for E1,}\\
&=\frac{\hbar}{2M c}\hat{e_i}\cdot(\vec{L}+g_n\vec{S})  ,   &\text{        for M1}.
\end{align*} 
Here $M$ is the nucleon mass, $\vec{r}$, $\vec{L}$ and $\vec{S}$ are the spatial distance, orbital and spin angular momentum operators. $g_n$ is the g-factor for the transition particle. 
The subscript in $\hat{H}_{i}$ denotes that only the i$^{th}$ component part of Eq.~(\ref{eq4}) is considered. The value  $\omega=\omega_f-\omega_i$ ($\omega_{mi}=\omega_m-\omega_i$) is the energy gap between the initial and the final (intermediate) state. The term $(1\longleftrightarrow2)$ has the same form as the first term but with 1 and 2 exchanged.  

Note that here the sum over $|m\rangle$ belongs to eigenstates of the system, so that it cannot overlap with the virtual state $|\tilde{v}\rangle$ with $(E_{\tilde{v}}-E_i)/\hbar=\omega_1$ or $\omega_2$. Eq.~(\ref{eq6}) has the same form of time-dependence as 1PA, it then leads to
\begin{equation}
|a^{[2]}(t)|^2=\frac{e^4 \mathcal{E}^2_1 \mathcal{E}^2_2}{16\hbar^4}|\mathcal{M}|^2\frac{\sin^2[(\omega-\omega_1-\omega_2)t/2]}{[(\omega-\omega_1-\omega_2)/2]^2}, \label{pr}
\end{equation}
with
\begin{equation}
\mathcal{M}=\sum_m\left[\frac{\langle f|\hat{H}_{2}|m\rangle\langle m|\hat{H}_{1}|i\rangle}{\omega_1-\omega_{mi}}+(1\longleftrightarrow2)\right]. \label{m}
\end{equation}
Using the same asymptotic form at $t\rightarrow \infty$ (as appeared in Fermi's golden rule) gives the following definition of transition rate
\begin{equation}
R_{2pa}=\frac{e^4 \mathcal{E}^2_1 \mathcal{E}^2_2}{16\hbar^4}|\mathcal{M}|^2 2\pi\delta_t(\omega-\omega_1-\omega_2), \label{prob}
\end{equation}
where $\delta_t$ has dimension of [time].

\section{Derivations of effective cross section}

With the line-shape function $g$ given by Eq.~(7) in the main text, the effective cross-section for an incoming $\gamma$-photon can be obtained by firstly replacing the $\delta$-function in Eq.~(\ref{prob}) by $g$ (with $\tilde{R}_{2pa}$ denoting this change), and then dividing $\tilde{R}_{2pa}$ by the flux multiplied by its velocity, i.e., $I_{\gamma} c$. This leads to
\begin{eqnarray}
\sigma^{2pa}_{eff}=\frac{\tilde{R}_{2pa}}{I_{\gamma}c}G
. \label{sig}
\end{eqnarray}
Using the relation $4\pi I_i\omega_i=\frac{\varepsilon_0 c}{2}|\mathbf{E}_i|^2$, where $I_i$ is the intensity of photons and $\mathbf{E}_i$ is the i$^{th}$ component of Eq.~(\ref{eq4}), one has
\begin{eqnarray}
\sigma^{2pa}_{eff}=(2\pi)^3\frac{e^4 \omega_1\omega_2}{c^2\hbar^4 
 \varepsilon_0^2} I_2|\mathcal{M}|^2 gG
, \label{2pa_eff}
\end{eqnarray}
where $\varepsilon_0$ is the vacuum permittivity.
According to Weisskopf estimates (which gives reasonable matrix elements within a factor $\sim10^2$ with respect to experiments), the matrix element $|\langle \hat{H_i}\rangle|$ in Eq.~(\ref{m}) has the following form\footnote{The index $\langle i|$, $\langle f|$ and $\langle m|$ are dropped as the detailed wavefunction no longer matters in Weisskopf estimates.} 
\begin{eqnarray}
|\langle\hat{H_{i}}\rangle|=\sqrt{B(typ,l_i)}, \label{ma}
\end{eqnarray}
with $B(typ,l_i)$ the reduced transition probabilities
\begin{eqnarray}
B(E,l_{i})&=&\frac{1}{4\pi}\left[\frac{3}{l_{i}+3}\right]^2 R^{2l_{i}}, \\
B(M,l_{i})&=&\frac{10}{\pi}\left[\frac{3}{l_{i}+3}\right]^2 R^{(2l_{i}-2)}(\frac{\hbar}{2Mc})^2.
\label{b}
\end{eqnarray}
For dipole transition $l_{i}=1$. Therefore $|\langle \hat{H}^{E1}_i\rangle|\approx R/4.7$, where the radius of nucleus $R\sim r_0 A^{1/3}$ with $r_0\approx1.2$ [fm] and $A$ the number of nucleons. $|\langle \hat{H}^{M1}_i\rangle|\approx \frac{\sqrt{10}}{4.7}\frac{\hbar}{Mc}$.

\section{Details of the conversion in the $\gamma$ spectrum}
The $\gamma$ spectrum given in Fig.~2 in the main text adopts data points converted from the brilliance listed in Refs.~[119-123]. 
Note that the brilliance, $B_r$, is given in unit [s$^{-1}$mm$^{-2}$mrad$^{-2}$($0.1\%$BW)$^{-1}$]. Meanwhile, the crucial quantity is the flux of the photons per bandwidth (BW) coming out from the initial area they were generated. Nevertheless, the conversion is straightforward, provided that the duration of the $\gamma$-photons and the actual angular spread of the beam (denoted as $\theta$ here) are given. Denoting $N_{\gamma}$ (in the unit [eV$^{-1}$cm$^{-2}$]) the areal density of photons accumulated within 1 eV interval at a given energy $E_{\gamma}$, one has
\begin{eqnarray}
N_{\gamma}=\frac{B_r\times\text{[duration in fs]}\times[\theta \text{ in mrad}]^2}{10^{15}\times[10^{-3}E_{\gamma} \text{ in eV}]}10^2.\label{br}
\end{eqnarray}
Note that the curve in Fig.~2 is produced by an exponential fit from the data of Ref.~[119] at $E_{\gamma}=1$, 10 and 100 MeV. 
However, only at $E_{\gamma}=1$ MeV, $\theta=0.7$ [rad] is given. We have therefore assumed that the same $\theta$ applies for other energies\footnote{Normally, $\theta$ would be lower for higher energies.}. Thus, the black curve shown in Fig.~2 could be subjected to an uncertainty up to 10 times for $E_{\gamma}\gtrsim 10$~MeV in the worst scenario.  Similarly, for Refs.~[120-123], the value of $\theta$ is adopted whenever directly available or by interpreting its value from relevant plots (i.e., those showing the angular divergence of the beam). 


\section{Practical scenarios of the HPLS-based $\gamma$- and isomer-production}

To enable our isomer pumping/depletion scheme, one of the prerequisites is a high-density $\gamma-$flash, which can be given by the interaction between high-power laser pulses and matter. Although laser wakefield acceleration~\cite{lwfa, esarey2009physics} based radiation schemes\,\cite{albert} such as bremsstrahlung and betatron radiation are specific alternatives, shooting a high-power laser pulse onto an over-dense target may be the preferred approach because of the higher yields of $\gamma-$photons per eV interval, which can reach an areal density $N_{\gamma}\sim \SI{e12}{eV^{-1}cm^{-2}}$ for $E_{\gamma}\approx1$~keV$\sim \SI{5}{MeV}$, as extracted very conservatively (Ref.[119] suggests $N_{\gamma}\sim \SI{e14}{eV^{-1}cm^{-2}}$ for the same interval) from Fig.~2 in the main text.

This mechanism starts with a laser pulse of a wavelength of $\lambda  \sim  \SI{1}{\upmu m}$ that is compressed through the chirped pulse amplification technique~\cite{cpa} into a time scale of $\sim \SI{25}{fs}$ to \SI{50}{fs}  to achieve intensities up to $\mathcal{P}\sim \SI{e21}{Wcm^{-2}}$ to  $\mathcal{P}\sim \SI{e23}{Wcm^{-2}}$.
Such an ultraintense infrared pulse then interacts with a solid target and ionizes its electrons, forming a plasma environment. Consequently, this laser-matter interaction accelerates part of the electrons to relativistic energies.

These pre-accelerated electrons then interact with the laser field, emitting the high energy radiation, typically via the non-linear Thomson or Compton scattering mechanisms. Multiple schemes for generating dense $\gamma-$photons exist and are cited as [119-123] in the main text.
In those schemes, the temporal contrast of the HPLS can significantly influence the properties of generated $\gamma-$flash. However, there are techniques to improve the contrast, as for instance the plasma mirror, although at the expense of the main pulse energy.    Ref.~[119] in the main text suggests a 10-fold reduction of the $\gamma-$flux if the laser power is reduced from \SI{10}{PW} to \SI{3}{PW}. In this scenario, the data represented in Fig.~2 of the main text can practically be rescaled accordingly.  

During the generation of $\gamma-$photons, particles such as electrons, low-energy (i.e., eV-level) photons, ions, and target debris are usually produced. Although charged particles could potentially contribute to isomer pumping/depletion (e.g., via Coulomb excitation) and might be desirable, they could also destroy a solid-type isomer target. The most straightforward way to sufficiently attenuate the population of eV-level photons and charged particles, block the debris, and prevent crucial damage to the isomer target is to place a solid filter between the target of laser-matter interaction and the isomer target. As illustrated in Fig.~1 in the main text, such a filter also serves as a plasma mirror to absorb/reflect/scatter the laser light.

A filter of about 1 mm thickness or more (e.g., 1 mm Tungsten) is sufficient to stop the high population of low-energy particles, which can cause damage to the isomer target. Such a filter would still have a negligible effect in absorbing the $\gamma-$flash. However, the inclusion of such a filter will increase the distance between the $\gamma-$photon source and isomer target; hence, it is sensible to minimize its thickness using, for instance, high-$Z$ materials.

Note that a crucial advantage of our scheme is that the $\omega-$photons coming from the supplied laser are not required to be aligned to the $\gamma-$photons for multi-photon absorption to occur, although the spatial overlap between $\gamma-$photons, $\omega-$photons, and the isomer target is of the essence. The beam quality of the $\gamma-$photons determines the achievable overlap. PIC simulations (Refs.~[119-123]) suggest that the majority of the $\gamma-$photons are generated from an initial area $\sim \SI{10}{\upmu m} \times \SI{10}{\upmu m}$ and exhibit an angular divergence $\theta\sim \SI{30}{\deg}$. Therefore, the flux would be reduced by a factor $\sim$ 2000 after propagating $\sim$ 1 mm.
\begin{figure}[t]
\includegraphics[width=0.47\textwidth,clip=true]{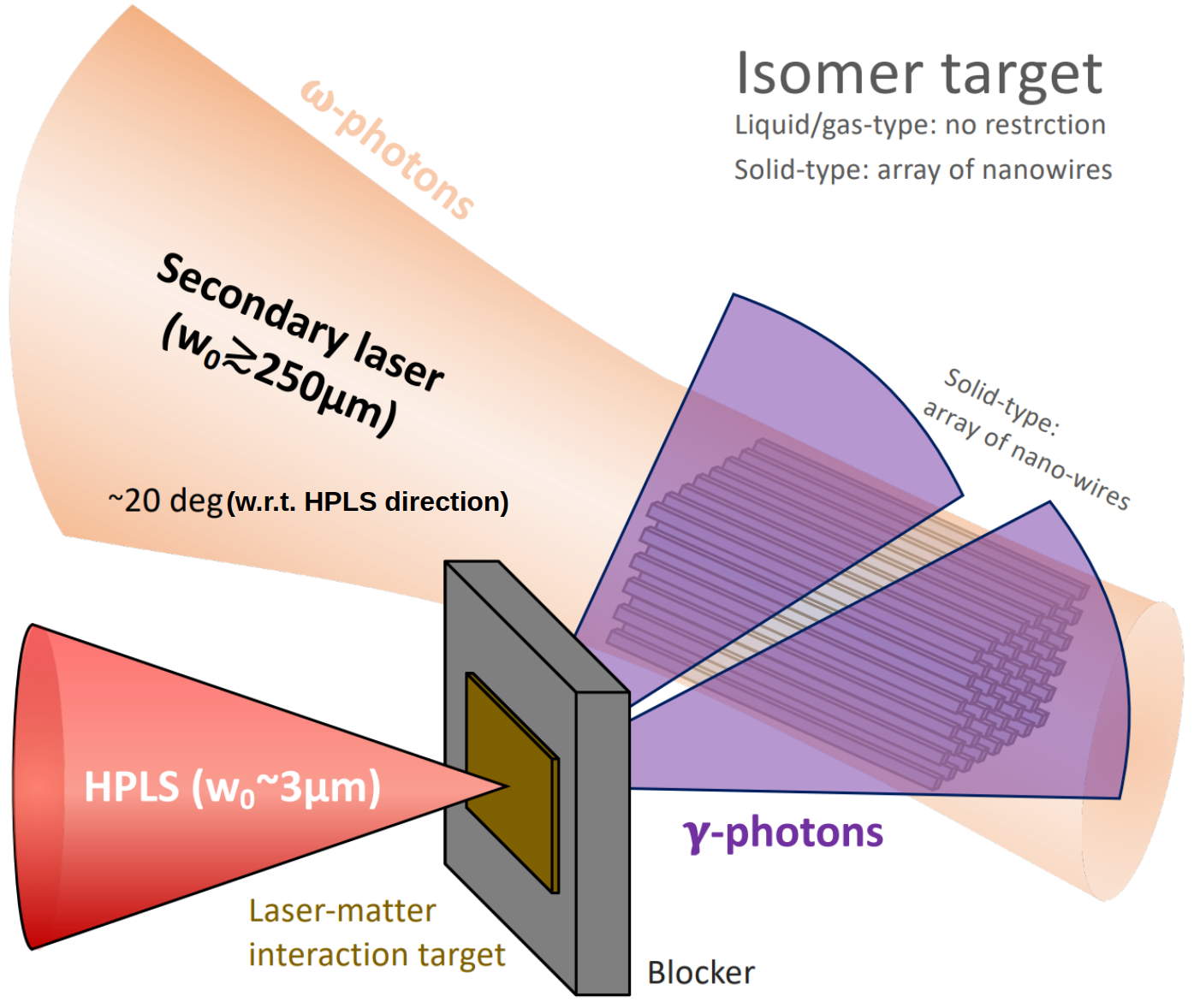}
\caption{Illustration of an experimental setup. A $\gamma-$flash driven by the main HPLS pulse typically has two main lobes, where the isomer targets are placed correspondingly. If the isomer target is of solid form, a nano-wire-like structure is preferred (with their detailed structure illustrated in Fig.~1 of the main text), and the $\omega-$photons need to be aligned to the nano-wires, which may be placed to have an angle $\sim\SI{20}{\deg}$ with respect to the HPLS direction. If the isomer target is of gas or liquid form, all restrictions can be levitated in favoring a maximum overlap volume between the $\gamma-$ and $\omega-$photons.}
\label{fig4_DD}
\end{figure} 
In any case, assuming a $\sim2000\times$ diluted $\gamma-$intensity is still remediable by the supplied $\omega-$photons\footnote{The divergence of $\gamma-$photons plays no role for a 2PA of $\omega+\gamma$ type, but it matters when more than one $\gamma$ enters the nPA.}, the supplied laser needs to be adjusted to have a waist $w_0\gtrsim \SI{250}{\upmu m}$ to cover the entire path of the $\gamma-$photons up to a distance $\approx$ 1 mm. This is very achievable even with TW-class laser systems and a long focal mirror to achieve a focused beam with a large focal spot and several mm of Rayleigh length. Owing to the waist size and Rayleigh length larger than the region of laser-matter interaction, the TW laser can be placed either on the side facing the laser-matter interaction (slightly non-parallel to the direction of HPLS, as illustrated in Fig.~\ref{fig4_DD} above), or on the opposite side---where it is allowed to propagate anti-parallelly with respect to the HPLS\footnote{In this case, the base where nano-wires are planted will be shifted to the HPLS side and integrate with the plasma mirror.}. 

One remaining factor in the practical realization concerns the practical target design to maximize the isomer yields. 
As mentioned in the main text, if eV-level $\omega-$photons are adopted, with $\mathcal{P}\gtrsim \SI{e12}{Wcm^{-2}}$, they can only penetrate solid targets up to a skin depth $\sim \SI{5}{nm}$. 
In this case, an array with a nano-wire-like structure for the isomer target is preferred. As illustrated in Fig.~1 in the main text, we estimate the dimension of each side of the nano-wire to be $\sim \SI{20}{nm}$, 
and an inter-wire spacing of $\gtrsim \SI{200}{nm}$. This will allow $\omega-$photons to propagate through the entire isomer target without being blocked or reflected by the nano-wires.

Finally, it is worth noting that the above setup and restrictions regarding the isomer pumping have relevance \textit{only} to a solid target and \textit{only} in the case where one wishes to keep its original shape in favor of pumping/depletion with continuous cycles. 
In general, gas or liquid forms of isomers or a one-time-use solid target can be employed with a much simpler experimental setup.

\bibliography{aipsamp}

\end{document}